\documentclass[smallextended]{svjour3}       

\smartqed  
\usepackage{graphicx}
\usepackage{amssymb}
\usepackage{amsmath}
\usepackage{amsfonts}

\usepackage{cases}
\usepackage{stfloats}
\usepackage{subfigure}
\usepackage{mdwlist}
\usepackage{threeparttable}
\usepackage[all,2cell]{xy} \UseAllTwocells
\usepackage{booktabs}
\usepackage{fancyvrb}
\usepackage{rotating}
\usepackage{rotfloat}
\usepackage{pifont}
\usepackage{verbatim}
\usepackage{lineno}
\usepackage{blkarray}
\usepackage{algorithmicx}
\usepackage{algorithmicx}
\usepackage[linesnumbered,ruled,vlined]{algorithm2e}
\usepackage[noend]{algpseudocode}
\usepackage{stfloats}
\usepackage{dsfont}
\usepackage{psfrag}
\usepackage{color,mdwlist}
\usepackage{subfigure}
\usepackage{bigstrut,array,multirow,tabularx}
\usepackage{bbm}

\usepackage{graphicx}
\usepackage{amssymb}
\usepackage{amsmath}
\usepackage{cite}
\usepackage{stfloats}
\usepackage{dsfont}
\usepackage{psfrag}
\usepackage{color,mdwlist}
\usepackage{subfigure}
\usepackage{balance}
\usepackage{lipsum}
\usepackage{mathtools}
\usepackage{cuted}
\usepackage{braket}
\usepackage[mathscr]{euscript}
\usepackage{calrsfs}
\usepackage{xcolor}
\usepackage{soul}
\DeclareMathAlphabet{\pazocal}{OMS}{zplm}{m}{n}

\DeclareMathOperator{\tr}{\mathrm{tr}}

\SetKwInput{KwInput}{Input}                
\SetKwInput{KwOutput}{Output}

\begin{document}

\title{ Adaptive quantum state tomography with iterative particle filtering 
}


\author{Syed Muhammad Kazim  \and  Ahmad Farooq    			\and
		Junaid ur Rehman			\and
        Hyundong Shin
}


\institute{
		Syed Muhammad Kazim \\
		kazim@khu.ac.kr \\~\\
		Ahmad Farooq \\
		ahmadfarooq@khu.ac.kr \\~\\
		Junaid ur Rehman \\
		junaid@khu.ac.kr \\~\\
		Hyundong Shin (Corresponding Author)\\
		hshin@khu.ac.kr \\~\\
		Department of Electronics and Information Convergence Engineering, Kyung Hee University, \\Yongin-si, 17104 Korea           
}
\date{Received: date / Accepted: date}

\maketitle


\begin{abstract}

Several Bayesian estimation based heuristics have been developed to perform quantum
state tomography (QST). Their ability to quantify uncertainties using region estimators
and include a priori knowledge of the experimentalists makes this family of methods an
attractive choice for QST. However, specialized techniques for pure states do not work
well for mixed states and vice versa. In this paper, we present an adaptive particle filter
(PF) based QST protocol which improves the scaling of fidelity compared to nonadaptive Bayesian schemes for arbitrary multi-qubit states. This is due to the protocol's unabating perseverance to find the states’ diagonal bases and more systematic handling of enduring problems in popular PF methods relating to the subjectivity of informative priors and the invalidity of particles produced by resamplers. Numerical examples and implementation on IBM quantum devices demonstrate improved performance for arbitrary quantum states and the application readiness of our proposed scheme.

\end{abstract}


\newpage

\section{Introduction} \label{sec:intro}

With increasing research and development in the fields of quantum information and computing, accurate identification and estimation of system components has become vital in efficient operation of practical quantum devices. Quantum state tomography (QST) is one such technique, which is used to characterize an unknown given quantum state $\rho$. More formally, QST statistically reconstructs a density matrix $\rho$ using data generated by measuring a large number $N$ of identically prepared copies of $\rho$.

Generally, the reliability and  accuracy of the estimated density matrix can be improved by increasing $N$. Due to the resource-intensive nature of QST, a key figure-of-merit of any QST technique is its scaling with respect to $N$, which can be improved by measuring $\rho$ is some optimal measurement basis. Therefore, special interest in QST has been given to the development of optimal basis for measurement \cite{PXB:11:PRA}, \cite{FMZ:10:PRA}. Improvement in measurement precision has been experimentally demonstrated when projectors of the eigenstates of Pauli operators employed in standard measurement strategies are supplanted by mutually unbiased bases (MUB) \cite{RA:10:PRL}. Despite the success of MUBs, it has been argued that static approaches utilizing a fixed set of measurements do not take advantage of the information obtained from measurements during QST \cite{HH:12:PRA}. Therefore, adaptive realizations of QST are better positioned to reduce redundancy.

It has been shown analytically that the worst-case infidelity can be reduced to $O \left(1/N\right)$ if $\rho$ is measured in its diagonal basis \cite{LLJ:18:PRA}. Since $\rho$ is unknown, its diagonal basis is also a priori unknwon. Adaptive techniques for QST have been developed in the past decade, which attempt to approach the diagonal basis of $\rho$ and employ it as a measurement basis. For example, popular maximum likelihood estimation (MLE) based adaptive schemes take a two-pronged approach: perform standard tomography on $N/2$ copies of $\rho$, estimate an intermediate state $\hat{\rho}$, and measure the remaining $N/2$ copies in the diagonal basis of $\hat{\rho}$ \cite{DLA:13:PRL}, \cite{ZHG:16:NPJ}. This two-stage tomography using MLE provides improved estimates of $\rho$ for any value of $N$ compared to other standard procedures. However, MLE has inherent issues as a statistical estimator for QST. Although it is possible to reconstruct states using MLE, its incompatibility with error bars makes it difficult to explain the uncertainty and thus the reliability of the estimates. Moreover, as MLE is primarily a frequentist construct, it fits observed frequencies obtained from essentially probabilistic measurements to probabilities \cite{KR:10:NJP}. In case of small data sets, MLE can be very unreliable.

Bayesian QST is an alternative to the popular MLE-based QST, which does not suffer from the fundamental drawbacks of MLE. In Bayesian QST, we augment our a priori knowledge of $\rho$ with a data driven likelihood function to produce a well-defined posterior distribution. The posterior distribution allows us to make statistical estimates about $\rho$ and evaluate quantifiable error bars \cite{KR:10:NJP, BL:18:Book, Fer:14:NJP}. Moreover, unlike the two-stage MLE, we do not need to know $N$ prior to our experiment.

Adaptive Bayesian schemes have been employed to good effect in QST. Adaptive Bayesian quantum tomography \cite{HH:12:PRA} employs a particle filter (PF) based approach, and uses an information-theoretic utility function which optimizes measurements in each iteration. Self-guided quantum tomography (SGQT) \cite{Fer:14:PRL} is another technique that uses optimization to solve QST for pure states, and has reported considerable improvement in the estimation of pure states. More recently, practical adaptive quantum tomography (PAQT) \cite{CCS:17:NJP}, a hybrid of PF and SGQT, has demonstrated that SGQT can be applied to mixed states with good effect. PAQT also reports improvements in fidelity over contemporary PF based QST \cite{HH:12:PRA} for states regardless of purity. Yet, SGQT still outperforms PAQT for pure states. Therefore, despite significant advancements in Bayesian QST, there is still a need for a single technique that is equally adept at estimation of both pure and mixed states, and provides scaling better than or similar to other formulations. That is, given a random qubit, we should not have to choose between different techniques based on, a difficult to justify, a priori assumption of the state's purity.

In this paper, we present a unified and adaptive practical technique, and report improvements in estimation over existing heuristics for random states of one, two and three qubits. Moreover, we also develop a state specific pseudo-prior such that the performance of QST is not incumbent on a priori insight of the experiment. Furthermore, we provide a resampling algorithm that corrects the propensity of creating invalid particles of popular resampling techniques. Lastly, we demonstrate the practical nature of the proposed scheme in the estimation of pure and mixed states by providing a proof-of-concept implementation on IBM's quantum computers \cite{GTP:20:Zen}.

This paper is structured as follows. We delve into the analysis of our prior, explain and provide a pseudocode for resampling of particles, and expound on the functionality of our adaptive protocol in Section~\ref{sec:2}. We exemplify our protocol in Section~\ref{sec:3}, and discuss and conclude in Sections~\ref{sec:4} and \ref{sec:5}, respectively.

\section{Methodology}\label{sec:2}
An arbitrary quantum state can be represented by a positive matrix of unit trace, i.e., a density matrix, commonly denoted by $\rho$. A general system of $n$ qubits, $\rho$ can be represented in terms of its Bloch vector $\pmb{r}$ \cite{MJW:01:PRA}
\begin{align}
	\rho = \frac{1}{d}\left(\mathbb{I} + \pmb{r}\cdot \pmb{\sigma}\right), 
\end{align}
where $d = 2^n$, $\mathbb{I}$ is the $d\times d$ identity matrix, $\pmb{r} = \left( r_1, r_2,\cdots, r_{d^{2}-1} \right) \in \mathbb{R}^{d^{2}-1}$ is a Bloch vector, and $\pmb{\sigma} = \left( \pmb{\sigma}_{1}, \pmb{\sigma}_{2}, \cdots\, \pmb{\sigma}_{d^{2}-1} \right)$ is a vector of Pauli words i.e. a tensor of two-dimensional Pauli operators. Normalization and positivity of $\rho$ translate to the norm constraints on the Bloch vector $\pmb{r}$, i.e., $\left\|\pmb{r} \right\|_{2} \leq 1$ where ${\left\| \cdot \right\|}_{2}$ is the Euclidean norm. Moreover, a sufficient description of valid quantum states requires $\rho$ to be positive semidefinite and hermitian with unit trace. Throughout this paper, we denote true state and the estimated state by $\rho$ and $\hat{\rho}$, respectively.

\subsection{Overview of PF-based QST}
The first common step for all QST schemes is to perform measurements on an ensemble of identically prepared copies of $\rho$. In the case of single qubits, when $\rho$ is measured in configuration $\alpha \in \pazocal{A}$ where $\pazocal{A}$ is a set of informationally complete projective measurement configurations, we observe one of the two possible outcomes $\ket{\psi_{\alpha}^{\ell}}$ for $ \ell \in \left\{ +1, -1 \right\}$. Let $N_{0}$ qubits be measured in the configuration $\alpha$, and $n_{\alpha}^{\ell}$ be the number of times we observe the outcome $\ell$. Then, the relative frequency $\hat{f}_{\alpha}^{\ell} = \frac{n_{\alpha}^{\ell}}{N_0}$ approximates the probability of outcome $\ket{\psi_{\alpha}^{\ell}}$ defined as $P\left( \ket{\psi_{\alpha}^{\ell}}\right) = \tr\left(\ket{\psi_{\alpha}^{\ell}}\bra{\psi_{\alpha}^{\ell}}\rho\right)$. Then what remains of QST is to best estimate the state $\hat{\rho}$ based on $\hat{f}$ using a statistical method that also specifies the uncertainty of the estimate.

In the QST implementation of Bayesian PF \cite{ASC:00:Book} \cite{LW:01:SMC}, we initialize particles $\left\{\gamma_k\right\}$ for $k \in \left\lbrace 1, \cdots, K \right\rbrace$ in the Bloch sphere that represent prospective states based on a priori knowledge or some other bias. A particle $\gamma_{k}$ is completely described at the $t$th iteration by its location $\pmb{r}_{k} \in \mathbb{R}^{d^{2}-1}$ and its weight $w_{k}^{t} \in \mathbb{R}$. The initial distribution of particles is known as the prior \cite{HH:12:PRA, CJD:16:NJP}
\begin{align}
	\text{Pr}\left(\pmb{r}\right) \approx \sum_{k} w_{k}^{0} \delta \left( \pmb{r} - \pmb{r}_{k} \right), 
\end{align}
where $w_{k}=\frac{1}{K}$. After performing $N_{0}$ measurements in $\alpha$ for the $t$th iteration, we calculate the likelihood \cite{DMC:12:QIC},
\begin{align}
\pazocal{L}\left( \gamma_{k} \mid \hat{f},\alpha \right) = N_{0}! \prod _{\ell} \frac{\tr \left( \text{p}_{a}^{\ell} \gamma_{k}\right)^{N_{0} \hat{f}_{a}^{\ell}}}{\left( N_{0} \hat{f}_{a}^{\ell} \right)!},
	\label{eq:loglike}
\end{align}
where $\text{p}_{a}^{\ell}$ is the projector on the eigenspace corresponding to the eigenvalue $\ell$ of $\sigma_{a}$ and update the weights of $\left\{ \gamma_{k}\right\}$ for the $(t+1)$th iteration as follows
\begin{align}
	w_{k}^{t+1} \approx w_{k}^{t} \times \pazocal{L} \left( \gamma_{k} \mid \hat{f},\alpha \right),
\end{align}
where $w_{k}^{t+1}$ is normalized such that $\sum_{k} w_{k}^{t+1}=1$. The resulting distribution $\left\{ w_{k}^{t+1}\right\}$ is the posterior for the $t$th iteration. Then the Bloch vector of $\hat{\rho}$ is the Bayesian mean estimate (BME) of the distribution, which is simply the weighted aggregate of the posterior
\begin{align}
	\pmb{r}_{\text{BME}} = \sum_{k} w_{k}^{t+1} \pmb{r}_{k}.
\end{align}
However, PF-based methods suffer from weight collapse where the whole posterior is concentrated on a single particle, giving it all the weight. This situation can be avoided by using an appropriate resampler that effectively reproduces the current distribution when the disparity in the weights of the particles exceeds a predetermined threshold.

In this section, we develop a prior $\text{Pr}\left(\pmb{r}\right)$, and identify the inefficiencies of contemporary resampling algorithms and introduce steps to resolve them. We also demonstrate the need and advantages of iterative learning of the set of measurement configurations $\pazocal{A}$ in adaptive Bayesian QST.

\subsection{Prior}

In Bayesian QST, informative priors \cite{BL:18:Book} can reduce the required number of copies $N$ of $\rho$ in the process. However, the quality of state estimation can suffer if the prior is based on incorrect insight. Although, attention has been afforded to increasing the robustness of protocols which rely on a priori knowledge, this robustness usually means the eventual convergence to the true state \cite{CJD:16:NJP}. That is, the variance in the required number of samples $N$ to achieve the same infidelity $\pazocal{I}$ for priors of varying authenticity of insight will be substantial. This is a significant problem since in real cases (state tomography of unknown states) where measurement metrics such as infidelity cannot be calculated to ascertain the accuracy of the process at any given stage, there is complete reliance on the statistical model to gauge infidelity for any value of $N$. Large variances in the output reduce our trust in the model, and hence reduces its practical applicability.

To counter these problems, we need a prior with statistically quantifiable errors. For this purpose, we utilize a small fraction of $N$ to obtain an initial rough estimate of $\rho$ and use a statistical model that specifies our region of interest in the Bloch sphere quantifying our uncertainty in the estimate. 

In our protocol, we use a Multivariate Gaussian distribution $\pazocal{N}\left(\hat{\pmb{\mu}}, \pmb{\Sigma} \right)$ with mean $\hat{\pmb{\mu}}$ and covariance $\pmb{\Sigma}$ as a prior for QST, which henceforth we refer to as our preliminary guess (PG). Initially, we assume no knowledge of the true state $\rho$. We prepare $\left( d^{2} - 1 \right) N_{0}$ copies of $\rho$ where $N_{0} \ll N$ and measure each Pauli word $\pmb{\sigma}_{j}$ for $j \in \left\lbrace 1,2, \cdots, d^{2}-1 \right\rbrace$ on $N_{0}$ copies of the state so that $n_{j}^{+}$ $\left( n_{j}^{-} \right)$ is the number of times we observe the outcome corresponding to the +1 (-1) eigenstate of $\pmb{\sigma}_{j}$.
Then $\pmb{\hat{\mu}} = \left( \hat{r}_{1}, \hat{r}_{2}, \cdots, \hat{r}_{d^{2}-1} \right)$ is the Bloch vector of the most likely state $\hat{\rho}$ after $\left( d^{2} - 1 \right) N_{0}$ measurements, where $\hat{r}_j = \frac{n_{j}^{+} - n_{j}^{-}}{n_{j}^{+} + n_{j}^{-}}$. In case $\left\| \hat{\pmb{\mu}} \right\|_{2} > 1$, we project it onto the surface of a $\left( d^{2} - 1 \right)$ dimensional ball by normalizing it. Note that ensuring unit norm is sufficient for $d=2$, but only necessary for $d>2$. Then for $d>2$, $\hat{\rho}$ may still not be a valid state, but we take no further actions at this stage.

To find the covariance matrix of PG, we must quantify our uncertainty in $\pmb{\hat{\mu}}$. That is, we must find the variance of the sample distribution of our estimate $\hat{r}_{j}$, or simply the square of the standard error of $r_{j}$ for the diagonal elements of $\pmb{\Sigma}$. First, note that the off-diagonal elements of $\pmb{\Sigma}$ are negligible because we perform projective measurements on Pauli words. Furthermore, since we perform $N_{0}$ measurements only, we can use an unbiased estimate of the variance to evaluate the standard error. Then, the $j$th diagonal entry 
$$\pmb{\Sigma}_{jj} = \frac{\sum_{i=1}^{N_{0}} \left( l_{i} - \hat{r}_{j} \right)^{2}}{N_{0}\left(N_{0}-1\right)} =
\frac{n_{j}^{+} \left( 1- \hat{r}_j \right)^{2} + n_{j}^{-} \left( -1- \hat{r}_j \right)^{2}}{N_{0}\left(N_{0}-1\right)} + \epsilon,
$$
where $\epsilon$ is a small constant defined to accommodate all $\rho$ that are eigenstates of our measurement operators. The $\left(d^{2}-1\right) \times \left(d^{2}-1\right)$ covariance matrix $\pmb{\Sigma}$ is a measure of the variances of the distances of $\hat{\rho}$ from $\rho$ after $N_{0}$ measurements in each element of $\pmb{\sigma}$. Furthermore, the diagonal elements of $\pmb{\Sigma}$ are state specific. For $N_0 = 50$, when $\hat{r}_j \rightarrow \left\{+1, -1 \right\}, \Sigma_{jj} \rightarrow 0$ and $\hat{r}_j \rightarrow 0, \Sigma_{jj} \rightarrow 0.02$. This state specificity is especially useful for states of high purity since it reduces our volume of interest for the same confidence level. Note that since we perform measurements to approximate parameters of $\pazocal{N}\left(\hat{\pmb{\mu}}, \pmb{\Sigma} \right)$ for PG, we can think of PG as a pseudo-prior rather than a conventional prior since the latter is constructed solely from a priori knowledge.

By utilizing only a fraction of copies, we have accomplished (i) an approximate state that serves as the first step for an adaptive QST technique explained later in the section, and (ii) reduced the region of interest substantially while retaining a statistical description of the uncertainty of PG.
To inspect (ii), let $\hat{\pmb{r}}= \left( \hat{r}_{x} , \hat{r}_{y}, \hat{r}_{z}\right)$ be the Bloch vector of a single qubit, and based on our previous discussion, let each element of $\hat{\pmb{r}}$ be an independent and identically distributed (i.i.d) Gaussian random variable with $\pazocal{N} \left(r_{i}, \Sigma_{ii}\right)$. An ellipsoid can be drawn to represent specific confidence levels using
\begin{align}
	\sum_{i \in \left\{x, y, z \right\}} \frac{{\left(\hat{r}_{i} - r_{i}\right)}^2}{\Sigma_{ii}} = s,
\end{align}
where $s$ represents the confidence interval of a $\chi^2$ distribution with three degrees of freedom and $s = 11.345$ for a $99\%$ confidence interval. In the case of a maximally mixed state, worst case for our strategy, the diagonal elements of $\pmb{\Sigma}$ are all equal to 0.02, and our ellipsoid corresponds to a sphere with diameter $2\sqrt{cs} = 0.135$. Therefore, compared to an uninformative prior which suggests all valid states are equiprobable, we have reduced the volume of interest by at least $99.97\%$. Therefore, as demonstrated in Fig.~\ref{fig:Prior}, by using a `just enough' informative prior which is characterized by its uncertainty works just as well as a good informative prior, and better than one characterized by inaccurate a priori knowledge.

\subsection{Resampling}

\begin{figure}[t]
	\centering
	\subfigure[]{
		\includegraphics[width=0.47\textwidth]{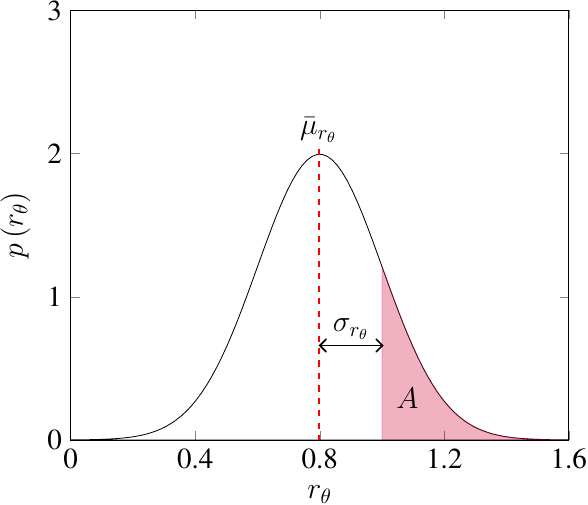}
	}
	\subfigure[]{
		\includegraphics[width=0.47\textwidth]{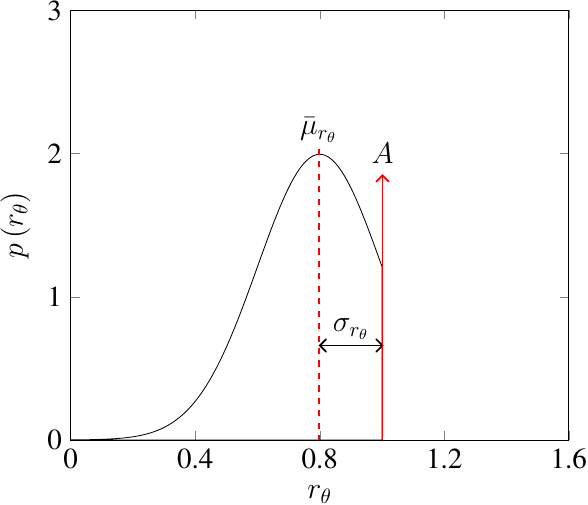}
	}
	\subfigure[]{
		\includegraphics[width=0.47\textwidth]{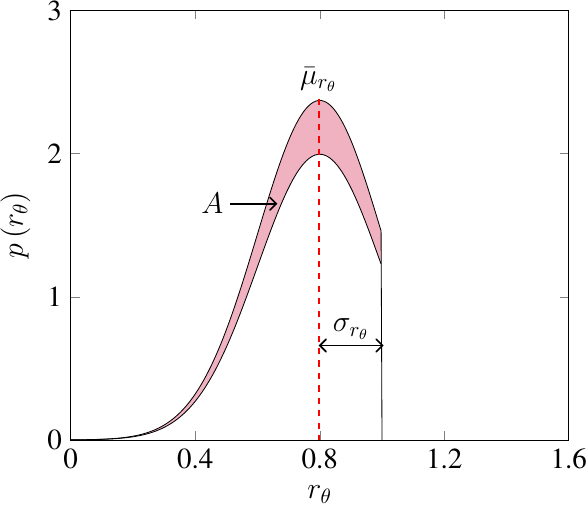}
	}
	\caption{
		1-D representation of Bloch sphere where $\theta \in \left\lbrace x, y, z\right\rbrace$ and $\left\|r_{\theta}\right\| \leq 1$. $\left(a\right)$ Resampler with PDF $\pazocal{N}\left(\bar{\mu}_{r_{\theta}}, \sigma_{r_{\theta}}\right)$. $\left(b\right)$ Output of resampler with concurrent negative eigenvalue truncation schemes morph the PDF such that there is an impulse at $r_{\theta}=1$ of size $A$. $\left(c\right)$ Proposed truncated Gaussian distribution such that the resampler outputs valid states without changing spatial probability ratios of samples.
	}
	\label{fig:resampler}
\end{figure}
In the PF-based Bayesian estimation, particles have to be resampled when $\sum_{k=1}^{K-1} w_{k}\approx 0$ and $w_{K}\approx 1$ \cite{PBT:08:book}. Depending on the resampling algorithm and proximity of the previous state to the surface of the Bloch sphere, resampled particles can be invalid states. Invalid two-dimensional states can be visualized as particles not circumscribed by the Bloch sphere, or more generally, as states represented by density matrices with eigenvalues less than zero or greater than one. A common workaround is to reduce the negative eigenvalues of such particles to 0, and normalize the remaining eigenvalues to produce valid states \cite{CJD:16:NJP}. Although this approach is simple, it deforms the resampling distribution, and increases computational expense since correction of thousands of particles can be required every time the particles are resampled. That is, when a high density of particles are close to the surface of the Bloch sphere (when estimating a pure state, or a state close by), a high number of samples produced after resampling will be invalid states, which are essentially projected onto the Bloch sphere as pure states. 

\alglanguage{pseudocode}
\begin{algorithm}[t]
	\small
	\caption{Practical resampling for valid states }
	\label{Algorithm:resampling}
	
	\KwInput{Particle weights $\left\lbrace w_{k} \right\rbrace$, locations $\left\lbrace \pmb{r}_{k} \right\rbrace$ for $k \in \left\lbrace 1, \cdots, K \right\rbrace$}
	\KwInput{$a \in \left[0,1\right]$}
	\KwOutput{Updated weights $\left\lbrace w_{k}' \right\rbrace$, locations $\left\lbrace \pmb{r}_{k}' \right\rbrace$}
	
	$\pmb{\mu} \leftarrow$ Mean($\left\lbrace w_{k} \right\rbrace$, $\left\lbrace \pmb{r}_{k} \right\rbrace$)\\
	$h \leftarrow \sqrt{1 - a^{2}}$\\
	$ \pmb{\mathrm{\Sigma}} \leftarrow h^{2}$ Cov($\left\lbrace w_{k} \right\rbrace$, $\left\lbrace \pmb{r}_{k} \right\rbrace$)\\
	$\left[\pmb{\mathrm{V}}, \mathrm{\pmb{\lambda}} \right] \leftarrow \mathrm{eig} \left(\pmb{\mathrm{\Sigma}}\right) \hspace{1cm} \blacktriangleright \pmb{\mathrm{V}}$ is a matrix of eigenvectors, $\mathrm{\pmb{\lambda}}$ is a vector of eigenvalues\\
	$\tau_{1} = \tau_{2} = \cdots = \tau_{d^{2}-1} = 0$\\
	
	\For{$k \in 1 \rightarrow K$}
	{
		draw $k^{th}$ particle $\pmb{r}_{k}$ with probability $w_{k}$\\
		$\pmb{{\mu}}' \leftarrow \pmb{\mathrm{V}}^{\text{T}}  \left( a\pmb{r}_{k} + \left(1-a\right)\pmb{\mu}\right) \hspace{3.22cm} \blacktriangleright {\pmb{\mu}}'= \left( {\mu}'_0, {\mu}'_1, \cdots, {\mu}'_{d^{2}-1} \right)$\\
		
		\For{$j \in \left\lbrace 1,2, \cdots, d^{2}-1 \right\rbrace$}
		{
			
			$C_{1} \leftarrow -\sqrt{1- \sum\limits_{l=0}^{j} \tau_{l}^{2}}$\\	
			$C_{2} \rightarrow -C_{1}$\\
			$\tau_{j} \leftarrow \text{TG}_{\left[C_{1}, C_{2}\right]} \left( {\mu}'_{j}, \sqrt{\mathrm{\lambda_{j}}}\right)$
		}
		$\pmb{r}_{k}' \leftarrow \pmb{\mathrm{V}} \cdot \left( \tau_{0}, \tau_{1}, \tau_{2} \right)^{\text{T}}$\\
		$w_{i}' \leftarrow 1/K$
	}
	return $\left\lbrace w_{k}' \right\rbrace \left\lbrace \pmb{r}_{k}' \right\rbrace$
	
\end{algorithm}

More specifically, suppose that we utilize a univariate Gaussian distribution $\pazocal{N}\left(\bar{\mu}_{r_{\theta}}, \sigma_{r_{\theta}}\right)$ to resample particles as shown in Fig.~\ref{fig:resampler}(a) where the shaded region $A$ represents the probability of sampling an invalid state. By projecting the invalid states onto the surface of the Bloch sphere, we have essentially sampled pure states with probability $A$ and used a distribution represented in Fig.~\ref{fig:resampler}(b). To accommodate the constraints of a two-dimensional valid density matrix without deforming the distribution of the particles and involving further computation required to correct invalid states, we resample using a truncated Gaussian (TG) distribution as shown in Fig.~\ref{fig:resampler}(c). To extend this approach to higher dimensional quantum states, we sample from a $\left( d^{2} - 1 \right)$ dimensional unit ball instead of the more complicated space of valid quantum states. Although this simplification precludes the ability of this algorithm to $\emph{always}$ sample valid states, it ensures smooth distributions for resampling. Procedure for sampling from the probability density function (PDF) of Fig.~\ref{fig:resampler}(c) is given in Algorithm~\ref{Algorithm:resampling}. For each particle, we sample $r_{k}$ for $k \in \left\lbrace1,2, \cdots, d^{2}-1 \right\rbrace$ from marginal Gaussian distributions corresponding to orthogonal axes, which are principal components of the current particle distribution, sequentially. We update the domain $\left[ C_{1} , C_{2} \right]$ of the next PDF based on the sample such that the constraint $\left\|\pmb{r}\right\| \leq 1$ is not violated.
Since we sample from marginal distributions of each axis independently, we must ensure that we can extract marginal PDFs from the available joint PDF. Therefore, we use $\pmb{\mathrm{V}}$ to change the system's bases to the principal components of the existing particle distribution. In this way, we remove existing correlations between bases. Lastly, for each orthogonal axis, we calculate $C_{1}$ and $C_{2}$ of Gaussian distribution, and sample $\tau$. $\pmb{r}_{k}'$ and $w_{k}'$ are the updated particle's location and weight, respectively.

\subsection{Adaptive Bayesian QST}

Infidelity scales as $O \left(1/N \right)$ when $\rho$ is measured in its own diagonal basis, which is the best possible scaling for QST. However, when this is not the case, it scales as $O \left(1/\sqrt{N} \right)$ for qubits close to the surface of the Bloch sphere \cite{LLJ:18:PRA},\cite{DLA:13:PRL}. Since $\rho$ is unknown, the purpose of introducing adaptive protocols is to approach the diagonal basis without losing information gained from intermediary measurement bases. Particle filtering naturally incorporates adaptivity in QST with minimal computational overhead.

\begin{figure}[t]
	\centering
	\subfigure[]{
		\includegraphics[width=0.4\textwidth]{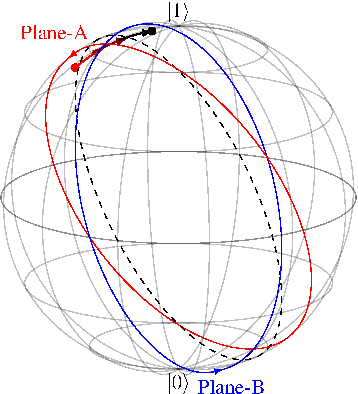}
	}
	\subfigure[]{
		\includegraphics[width=0.4\textwidth]{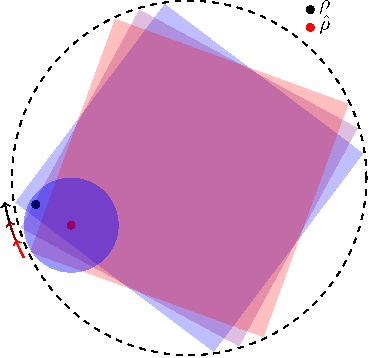}
	}
	\caption{
		$\left(a\right)$ Rotation of planes inside Bloch sphere based on estimated $\hat{\rho}$ where on average the plane move towards $\rho$ with increasing number of iterations. $\left(b\right)$ Opposite vertices of inscribed squares represent eigenstates of measurement operators where the adaptive scheme rotates the square such that it moves closer and eventually encompasses the true state to improve scaling of QST.}
	\label{fig:SPA}
\end{figure}

The formation of PG is conceptually the first step of our adaptive QST protocol, in that we initially perform measurements on $\rho$ using Pauli operators, and estimate a state, $\hat{\rho}$. We then proceed to change the measurement configuration for the next set of measurements. We use the eigenbases of $\hat{\rho}$ to rotate Pauli words such that the updated operators $\Omega_1$, $\Omega_2, \cdots, \Omega_{d^{2}-1}$ maintain $\text{Tr}\left( \Omega_{i}^{\dagger} \Omega_{j} \right) = 2\delta_{ij}$ and one of the operators diagonalizes $\hat{\rho}$. The particle filter updates the existing distribution based on the outcomes of the measurements in this configuration. We estimate an updated $\hat{\rho}$ and repeat the process iteratively until we have exhausted $N$ or some other criterion is met.

This iterative process is analogous to rotating the plane of measurement (in a two-dimensional space) and cube (in a three-dimensional space) such that the true state eventually lies within it, as shown in Fig.~\ref{fig:SPA}, to achieve the best possible scaling of infidelity. The circular plane represented by a dashed line on the left in Fig.~\ref{fig:SPA} is observed in the same figure on the right. The opposite corners of the squares circumscribed by the circle represent the eigenstates of the measurement operators. At a certain step in QST, state $\hat{\rho}$ is approximated. The plane of measurement is rotated such that eigenbases of one of the operators diagonalizes $\hat{\rho}$. The next set of measurements are performed in this configuration, and based on the outcomes a new plane of measurement is selected. The iterative rotation is indicated by the color of the planes in the direction corresponding to the arrows. States close to the surface of the Bloch sphere require a greater number of iterations. In the proposed scheme, we perform $N_0$ measurements on $\rho$ using a single operator $\Omega_{j}$ per iteration, where $\Omega_{j}$ is chosen randomly.
\begin{figure}[t]
	\centering
	\subfigure{
		\includegraphics[width=1\textwidth]{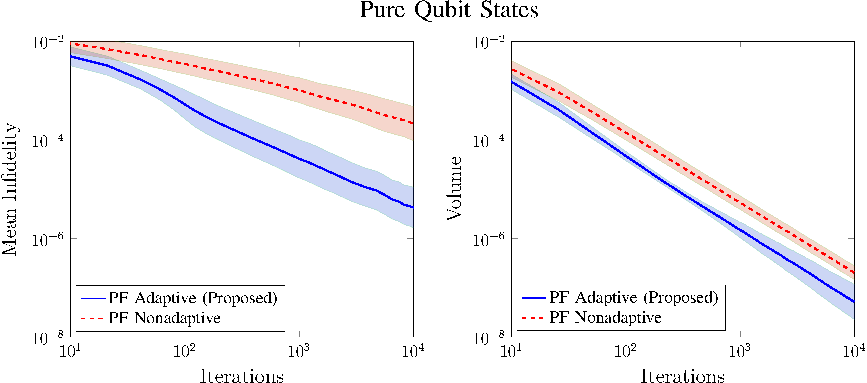}
	}
	\subfigure{
		\includegraphics[width=1\textwidth]{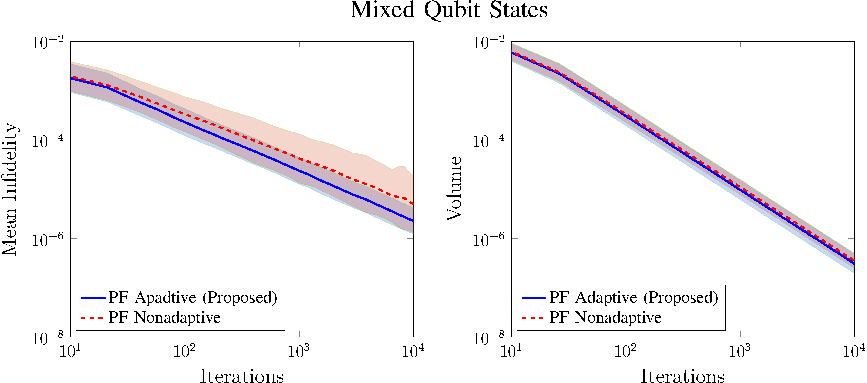}
	}
	\caption{
		Infidelity and minimum volume enclosing ellipsoid (MVEE) for $99\%$ credible region for the proposed adaptive protocol and its non-adaptive version for pure and mixed states using the method detailed in Section~\ref{sec:2} averaged over 1000 states. The shaded region indicates the $16\%$ and $84\%$ quantiles over all measurements.
	}
	\label{fig:Infid_vol_sim}
\end{figure}
\begin{figure}[t]
	\centering
	{
		\includegraphics[width=0.55\textwidth]{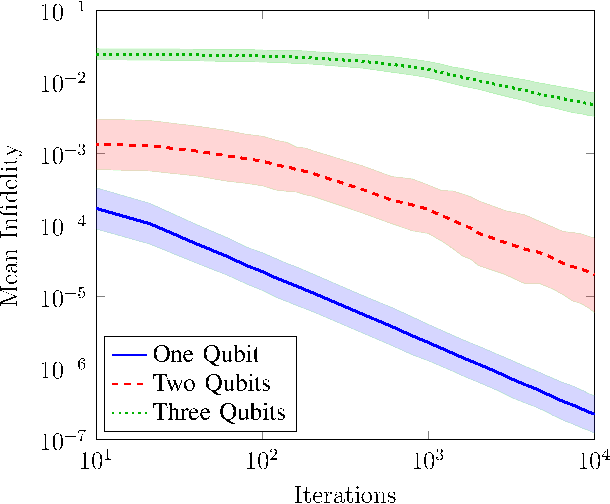}
	}
	\caption{
		Infidelity of one, two and three qubit mixed states. One qubit numerical results used 500 shots per operator for PG, whereas both two and three qubit cases used 2500 shots per operator. Each iteration consists of 500 shots. For one, two and three qubits, we used 2000, 4000 and 8000 particles, respectively. One and two qubit cases are averaged over 500 randomly sampled states from Hilbert-Schmidt uniform distribution whereas three qubit case is averaged over 50 states from the same distribution. The shaded area corresponds to $\pm 1$ standard deviation.
	}
	\label{fig:high_dim}
\end{figure}

\section{Numerical and Experimental Results}\label{sec:3}

In this section, we demonstrate the performance of our protocol detailed in Section~\ref{sec:2}. We perform QST using open-source packages Qinfer \cite{CCI:17:Qua} and Qutip \cite{JPF:12:CPC} on pure and mixed states using our adaptive protocol, and compare the results to non-adaptive PF algorithm. Moreover, we also perform QST on IBM quantum experience \cite{GTP:20:Zen} to show the readily applicable nature of our work. QST is performed on $\rho$ which is randomly sampled from Hilbert-Schmidt uniform and Haar uniform distributions for mixed and pure states, respectively, with the resampling parameter $\alpha = 0.1$. We report infidelity $\pazocal{I}$ between the estimated states $\hat{\rho}$ and true states $\rho$ defined as \cite{NC:10:Book}
\begin{align}
	\pazocal{I}\left(\rho, \hat{\rho}\right) =1 - \tr\left(\sqrt{ \sqrt{\rho}\hat{\rho}\sqrt{\rho}}\right),
	\label{eq:infid}
\end{align}
where fidelity $\pazocal{F} = \tr\left(\sqrt{ \sqrt{\rho}\hat{\rho}\sqrt{\rho}}\right)$ so that $
\pazocal{I} = 1 - \pazocal{F}$.
The infidelity captures the idea of closeness between $\rho$ and $\hat{\rho}$ such that $\pazocal{I}\left(\rho, \hat{\rho}\right) = 0$ if and only if $\rho = \hat{\rho}$. Note that $\hat{\rho}$ can be very close to $\rho$ and still be an invalid state due to our simplification of the valid space in Section~\ref{sec:2}.3. Therefore, as a final step, we check for the validity of $\hat{\rho}$, and ensure semi-definiteness by reducing negative eigenvalues to zero, and normalizing the remaining to ensure unit trace, when required. We also report volumes of covariance $\pmb{\Sigma}$-based ellipsoids enclosing 99\% credible regions defined as \cite{Fer:14:NJP}, \cite{CCN:12:NJP}
\begin{align}
	\text{Vol}\left(\pmb{\Sigma}\right) = \frac{\pi^{\left( d-1 \right) / 2}}{\Gamma\left(\frac{d}{2}+1\right)}\text{det}\left(\pmb{\Sigma}\right)^{-1/2},
	\label{eq:vol}
\end{align}
where $d$ is the dimension of ${\pmb{\Sigma}}$ and $\Gamma$ is the Gamma function for our simulations and experiments. This measure specifies the concentration of particles by calculating the volume enclosed by a fixed percentage of particles. Thereby helping us gauge the quality of convergence in successive iterations.

\begin{figure}[t]
 \centering
 \subfigure[]{
	\includegraphics[width=0.47\textwidth]{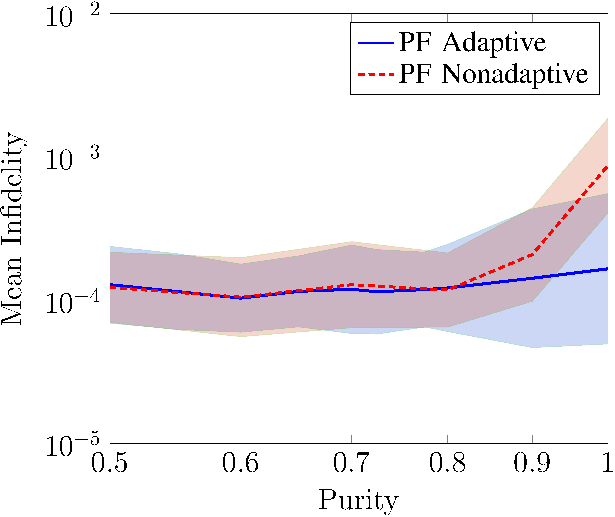}
	}
	\subfigure[]{
	\includegraphics[width=0.47\textwidth]{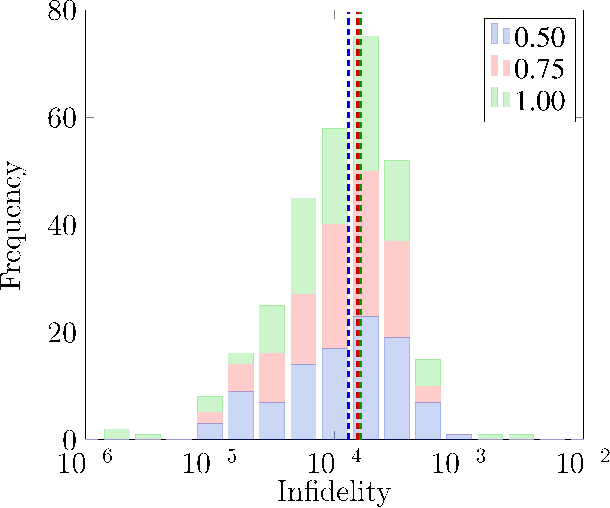}
	}
    \caption{
Invariance under purity of target state. (a) Mean infidelity of the proposed adaptive protocol and its nonadaptive version after $10^{4}$ measurements shown as a function of purity of the two-dimensional state to be estimated, averaged over 100 states. The shaded region corresponds to $\pm 1$ standard deviation. (b) Stacked histogram for infidelity using adaptive protocol of 100 states with purity 0.5 (blue), 0.75 (red) and 1.00 (green) each with their corresponding means given by dashed lines of the same colours, respectively.
}
    \label{fig:purity}
\end{figure}

Fig.~\ref{fig:Infid_vol_sim} demonstrates the advantage of proposed adaptive scheme as compared to non-adaptive PF-based scheme, both in terms of infidelity and the volume for 99\% credible regions. The difference between the two schemes is more pronounced for pure states because of the presence of only a single eigenvalue, which reduces uncertainty in measurement in an adaptive setting when the measurement operator diagonalizes a $\hat{\rho}$ that is close to $\rho$. Fig.~\ref{fig:high_dim} demonstrates the infidelities of one, two and three qubit mixed states.

Fig.~\ref{fig:purity} demonstrates the performance of the protocol as a function of the purity, given by $\tr \left(\rho^{2}\right)$. For each value of purity, 100 states were randomly sampled from the Hilbert-Schmidt uniform distribution. Then given a specific purity and keeping the eigenvectors of the sampled state unchanged, eigenvalues were updated accordingly. Fig.~\ref{fig:purity}(a) shows that performance of the adaptive protocol remains mostly invariant with respect to purity. Contrarily for the nonadaptive protocol, the performance declines as states become more pure. Fig.~\ref{fig:purity}(b) shows the distribution of infidelity of the states after $10^4$ measurements using the adaptive protocol. The distributions are very similar showing that the proposed protocol is unhindered by the state's purity.

\begin{figure}[t]
	\centering
	{
		\includegraphics[width=0.9\textwidth]{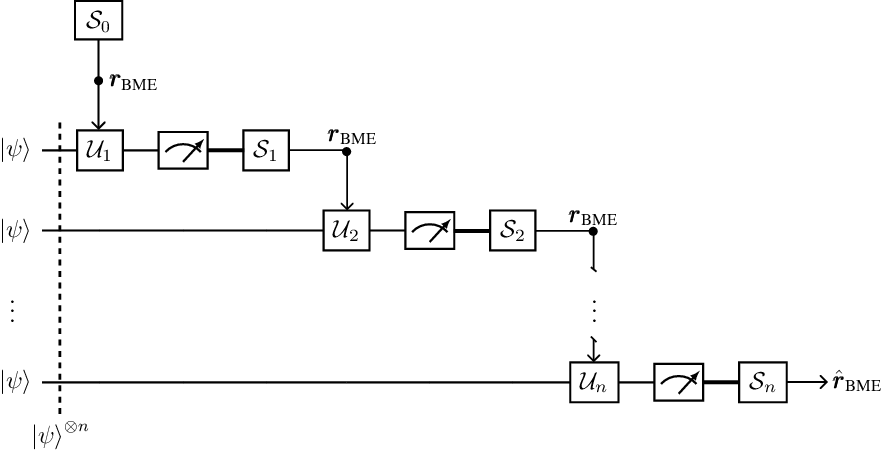}
	}
	\caption{
		The PF Adaptive QST for single-qubit state $\ket{\psi}$. 
		The $\pazocal{U}_{i+1}$ gate is configured based on the $\pmb{r}_{\text{BME}}$ of PF distribution $\pazocal{S}_{i}$ at iteration $i$. $\pazocal{U}_{i+1}$ changes the basis of measurement, and $\pazocal{S}_{i+1}$ is updated based on the measurements counts. The process is initialized with PG, $\pazocal{S}_{0} = \pazocal{N}\left( \pmb{\mu}, \pmb{\Sigma} \right)$, and $\hat{\pmb{r}}_{\text{BME}}$ is the Bloch vector of our estimate.
	}
	\label{fig:circuit}
\end{figure}

\begin{figure}[th]
	\centering
	{
		\includegraphics[width=0.55\textwidth]{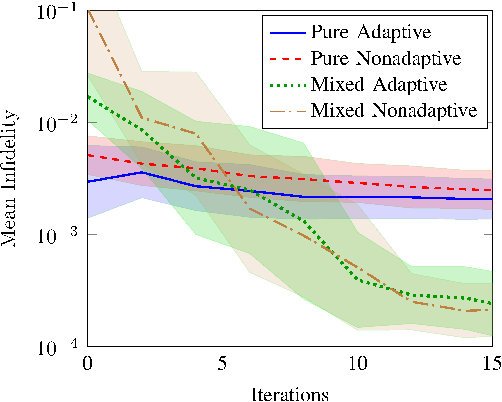}
	}
	\caption{
		Mean infidelity of pure and mixed qubit states from measurements on IBM quantum computers for the proposed scheme and its nonadaptive version, averaged over 50 states. Each iteration consists of $10^{3}$ shots where fidelity at $0\text{th}$ iteration is due to PG which performs three iterations of $10^{3}$ shots along each Pauli axis. The shaded area corresponds to $\pm 1$ standard deviation.
	}
	\label{fig:IBM}
\end{figure}

To perform QST on a mixed state on IBM's quantum computer \cite{IBM:20:BE}, we first prepare a two-qubit pure state $\ket{\psi} = \left[ 0.73, -0.25, 0.528, 0.348 \right] \in H_{B} \otimes H_{A}$ so that by taking the partial trace of $\ket{\psi} \bra{\psi}$ with respect to Hilbert space $H_{B}$, we attain $\rho \in H_{A}$ defined by the density matrix $\rho = \frac{1}{2} \left( \mathbb{I} + 0.6 \sigma_{x} + 0.2 \sigma_{z} \right)$, where $\sigma_x$ and $\sigma_z$ are Pauli X and Z respectively. At any specific iteration, we calculate the weighted aggregate $\pmb{r}_{\text{BME}}$ of the particle distribution to estimate $\hat{\rho}$, and rotate our measurement operators using the unitary operator $\pazocal{U}$ as detailed in Section~\ref{sec:2} and demonstrated by the quantum circuit flow diagram in Fig.~\ref{fig:circuit}. We measure the first qubit and update our particle filter accordingly. We also execute this process for a pure state, and report infidelity $\pazocal{I}$ for both states in Fig.~\ref{fig:IBM} for 15 iterations of 1000 shots each. In this paper we used \emph{ibmq\_athens}, which is one of the IBM Quantum Falcon Processors with average readout error and average CNOT error of $1.950 e^{-2}$ and $1.064 e^{-2}$, respectively.

In Fig.~\ref{fig:IBM}, the infidelity in the experimental implementation, scales better for mixed states than pure states. This difference is explained by state preparation errors of real quantum devices.
Due to the presence of noise, instead of preparing the state $\rho$, the quantum device prepares a different random state $\rho_{1}$ in each iteration, where $\rho_{1} \approx \rho$.
Let's assume that $\pmb{\sigma}$ diagonalizes $\rho$ and $\rho_{1}$. If $\rho$ is pure, then $1 - \epsilon \leq \tr \left(\rho_{1} \pmb{\sigma} \right) \leq 1$ for $\epsilon > 0$. This is why we also observe the early saturation of pure curves in Fig.~\ref{fig:IBM}. On the other hand, assuming state preparation errors to be inherently random, if $\rho$ is mixed, then $\tr \left( \rho \pmb{\sigma} \right) - \epsilon \leq \tr \left(\rho_{1} \pmb{\sigma} \right) \leq \tr \left( \rho \pmb{\sigma} \right) + \epsilon$. Since the expectation of the prepared states can be higher or lower than that of $\rho$, when $\rho_{1}$ is prepared independently in each iteration, the state preparation errors introduced in one iteration can be offset by other. Moreover,  improvement in later iterations is small, which is a typical behavior of tomography experiments on noisy systems \cite{GISK:16:PRA}. To obtain these results, we also utilized the noise mitigation offered by QISKIT module to reduce noise from quantum circuits and measurements that increased the number of measurements by a polynomial factor.
\section{Discussion}\label{sec:4}

\begin{figure}[t]
 \centering
 \subfigure[Pure States]{
	\includegraphics[width=0.47\textwidth]{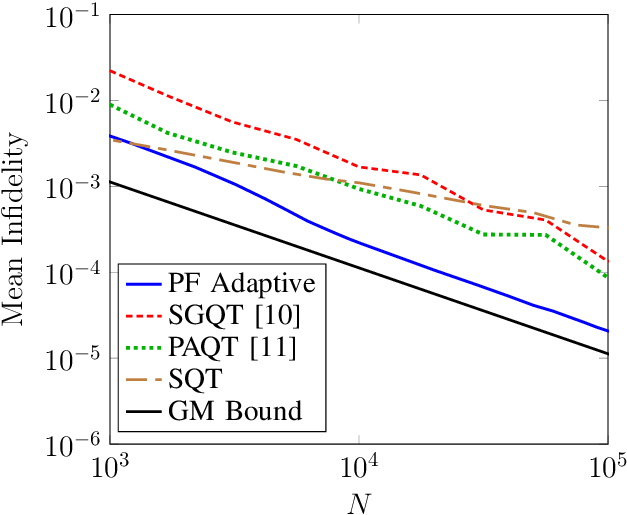}
	}
	\subfigure[Mixed States]{
	\includegraphics[width=0.47\textwidth]{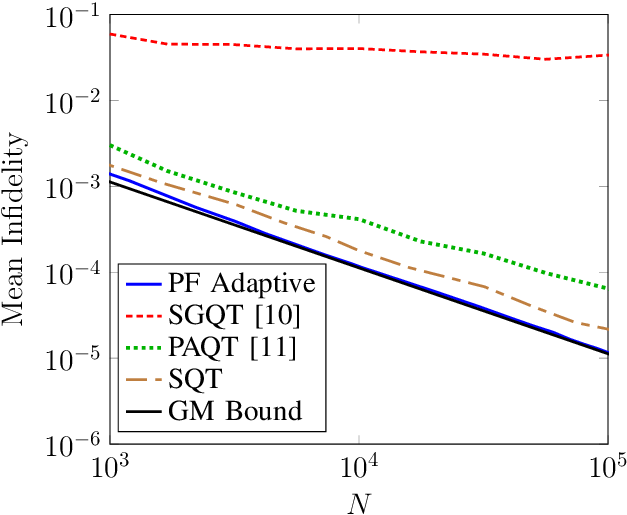}
	}
    \caption{
Comparison of mean infidelity of pure and mixed state estimations based on proposed protocol, PF Adaptive and its nonadaptive version, PF NA (averaged over 1000 states with 2000 particles) along with SGQT, PAQT (4000 particles) \cite{GFF:16:SUP} and the Gill-Massar (GM) bound. Each iteration consists of 50 shots along a measurement axis defined by the protocols.
}
    \label{fig:Infid_bay}
\end{figure}

Given that we set out to demonstrate that our method is adept at QST regardless of the purity of the state, we now show (in addition to Fig.~\ref{fig:purity}) its advantage over other contemporary Bayesian methods. Self-guided quantum tomography (SGQT) is a method that learns pure states through `Simultaneous Perturbation Stochastic Approximation' (SPSA) \cite{Fer:14:PRL}. However, it does not report error bars and works poorly for mixed states. Practical adaptive quantum tomography (PAQT) \cite{CCS:17:NJP} is a more rounded technique that builds on SGQT by applying measurements learned by SGQT on a Bayesian particle filter. PAQT can therefore report error bars, and estimate both pure and mixed states. In the process of making a unified technique, PAQT compromises on the infidelities reported by SGQT for pure states. In Fig.~\ref{fig:Infid_bay}, we demonstrate the mean infidelities of SGQT, PAQT and our method for both one qubit pure and mixed states against the number of measurements $N$. Data used for SGQT and PAQT plots is provided by the authors in \cite{GFF:16:SUP}. Our advantage over both techniques is visible especially when we compare there performance for states of arbitrary purity. 

The accuracy of estimated quantum state can be studied by means of relevant inequalities which are Cram\'er-Rao inequality $C \leq F^{-1}$, quantum Cram\'er-Rao inequality $C \leq J^{-1}$ and the Gill-Massar (GM) ineqaulity $\tr\left(F J^{-1}\right)$ $\leq d-1$, where $C, F$ and $J$ are covarinace, classical Fisher information and quantum Fisher information matrices, respectively \cite{GM:00:PRA}.
These fundamental inequalities established the lower bounds	 for several accuracy metrics, such as mean square error and infidelity, considering the impact of the finite ensemble size on the estimation uncertainty. The GM bound for the mean squared Bures distance in $d$ dimensional quantum system for finite number of copies $N$ is given as \cite{ZHU:12:THE}
\begin{align}
	D_{B}^{2}=\frac{1}{4N}\left(d+1\right)^{2}\left(d-1\right),
\end{align}
which can be defined through fidelity
\begin{align}
	1-\pazocal{F}^{2}\left(\rho,\hat{\rho}\right)=D_{B}^{2}-D_{B}^{4}/4.
\end{align}
Mean infidelity of the proposed protocol in Fig.~\ref{fig:Infid_bay} saturates the GM bound only for mixed single qubits. 
Moreover, for two or more qubits, infidelities obtained do not reach the GM bound. The diminished advantage for $d>2$ is explained by the increasing number of redundant iterations for larger systems. In the case of three qubits, after a sufficient number of iterations, only a single operator will produce outcomes which will be useful for state characterization, while the remaining 62 will produce approximately $\frac{N_0}{2}$ counts where $N_{0}$ measurements are performed in each iteration. Note that the GM bound in Fig.~\ref{fig:Infid_bay} is adjusted according to the definition of $\pazocal{I}$ provided in equation~\ref{eq:infid}.
\begin{figure}[t]
 \centering{
	\includegraphics[width=0.53\textwidth]{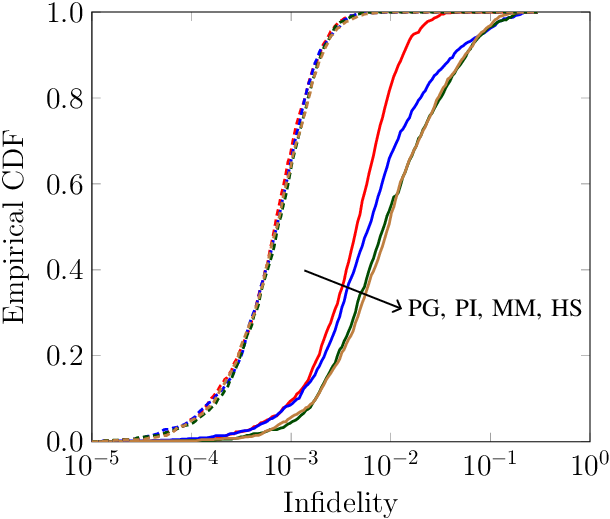}
	}
    \caption{
CDF of infidelity of 1000 randomly sampled mixed qubits. The initial guesses of partial information (PI), maximally mixed (MM) and Hilbert-Schmidt (HS) based priors for our QST protocol are $\hat{\rho}_0 = \left(1-p \right)\rho + \pi p$, $\frac{\mathbb{I}}{2}$, and randomly chosen state from HS distribution, respectively, where $p=0.1$, and the initial PF distribution is uniform. PG utilized $3 N_0$ samples prior to the first iteration, after which each iteration measures $N_0$ samples for all priors. For a fair comparison, $1\text{st}$ (solid) and $23\text{rd}$ (dashed) iterations of PG are plotted with $4\text{th}$ (solid) and $26\text{th}$ (dashed) iterations of PI, MM and HS respectively.
}
    \label{fig:Prior}
\end{figure}

Fig.~\ref{fig:Prior} compares the cumulative density function (CDF) of infidelity of our complete proposed scheme in which we utilize PG detailed in Section~\ref{sec:2} and three modified versions of our scheme which utilize priors of varying degrees of information. Partial information (PI), maximally mixed (MM) and Hilbert-Schmidt (HS) based priors are named to allude to the process with which we have estimated the initial $\hat{\rho}_{0}$ required to kickstart the adaptive QST protocol. So for PI, $\hat{\rho}_0 = \left(1- p \right)\rho + \pi p$ where $p=0.1$. For MM, $\hat{\rho}_0 = \mathbb{I}/d$ and for HS, $\hat{\rho}_0$ is a random sample from the Hiblert-Schmidt uniform distribution. However, since we are unsure of our uncertainty in $\hat{\rho}_0$, we have utilized a uniform particle filter distribution for each.
\begin{figure}[t]
 \centering{
	\includegraphics[width=0.58\textwidth]{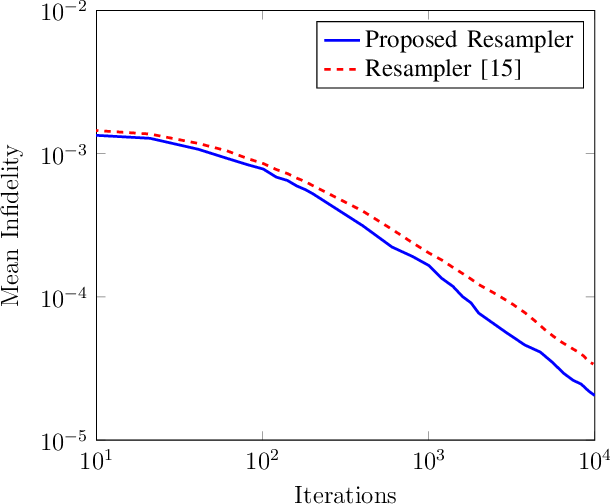}
	}
    \caption{
Mean infidelity of the proposed protocol with truncated Gaussian resampler and the conventional Liu-West resampler, averaged over 250 random states. Resampling parameter $\alpha=0.1$ and 4000 particles are used for state tomography of four-dimensional mixed states.
}
    \label{fig:resampler-2}
\end{figure}

While strictly speaking, PG is not a prior, it provides improved characterization of the quantum states initially. This improvement can be attributed to the ``free flowing" rotation of measurement operators and the concentration of particles in the proximity of the true state. In the case of PI, note that $\hat{\rho}$ shares the same eigenvectors as $\rho$. Therefore, in the first iteration, one measurement operator diagnoalizes $\rho$. This operator is chosen with probability $\frac{1}{ d^{2} -1 }$.
If this operator is chosen, then the probabilities of particles close to $\rho$ will increase in the posterior. However, since $N$ in small ($N=50$ is used for $d=2$ in this paper) and the particles are uniformly distributed, the cumulative probability of particles farther away decrease the proximity of the BME from $\rho$. This BME will be used to rotate the set of measurement operators in the next iteration, and therefore no operator will perfectly diagnolize $\rho$. In the case of MM, the operators will not rotate in the first iteration. After the first measurement the likelihood of the particles near $\tr{ \left( \Omega \rho \right)}$ will increase where $\Omega \in \left\lbrace X, Y, Z \right\rbrace$. Due to the spread of the particles, the operators will rotate in the next iteration but this rotation will be restricted. Lastly in the case of HS, $\hat{\rho}$ is just a randomly sampled state from the Hilbert-Schmidt uniform distribution, and such prior information cannot be expected to offer any improvement.
In all three cases, the advantage reduces with a higher number of iterations (once the resampler is called and rotations become more ``free") as shown by the overlapping CDFs (dashed) in Fig.~\ref{fig:Prior}. 

Fig.~\ref{fig:resampler-2} demonstrates the advantage of the proposed resampler over conventional resamplers \cite{LW:01:SMC} used in PF implementations in QST. For this figure, only the resampler is changed without making any other amendments to the adaptive scheme. Fig.~\ref{fig:resampler} demonstrated that conventional resampling strategies inadvertently allocate substantial probability to the edges of the Bloch space. Therefore, a larger number of particles are required to ensure they cover the required Bloch space. The problem exacerbates for higher dimensions.

For our protocol, we used  multiple values of the resampling parameter and found $\alpha = 0.1$ to be optimal. The intuition behind this is that on average the BME of the distribution at any iteration is closer to $\rho$ than any individual particle. Therefore, when resampling, we allow the mean of the resampler's distribution to be more heavily influenced by the BME than by the individual particles.
\begin{figure}[t]
 \centering
 \subfigure[Pure States]{
	\includegraphics[width=0.47\textwidth]{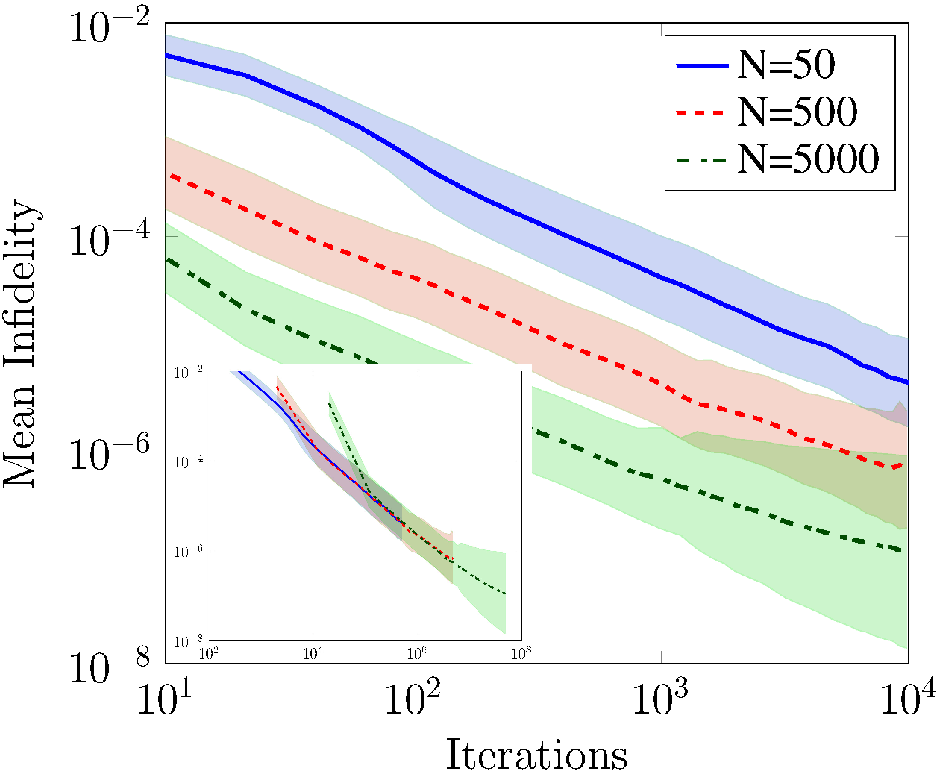}
	}
	\subfigure[Mixed States]{
	\includegraphics[width=0.47\textwidth]{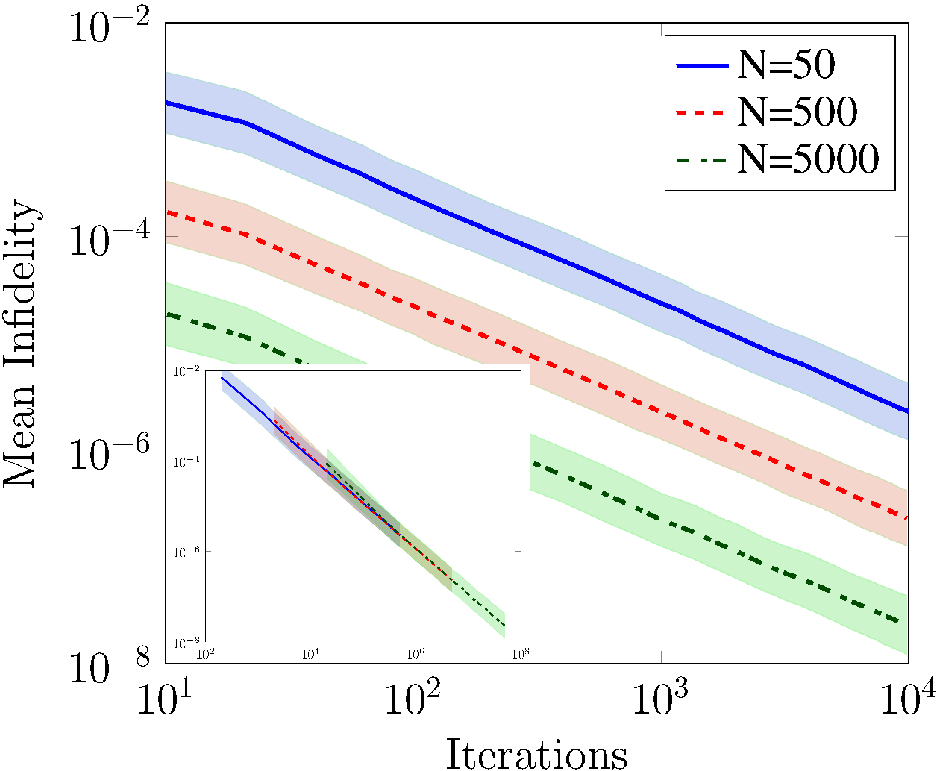}
	}
    \caption{
Comparison of mean infidelities of the proposed adaptive protocol for pure and mixed qubit states for different $N$, where $N$ is the number of shots along a measurement axis in each iteration. Each curve is averaged over 1000 randomly sampled particles and the shaded region indicates the $16\%$ and $84\%$ quantiles over all measurements.
}
    \label{fig:Noise}
\end{figure}

The conventional resampler in Fig.~\ref{fig:resampler-2} uses a Gaussian distribution to resample particles and then performs eigenvalue correction on each particle. In case we do not perform any correction to the sampled particles and considering that we sample a large number of particles (4000 for Fig.~\ref{fig:resampler-2}) multiple times for QST of a single state, it is highly likely that we will sample a significant percentage of invalid particles. Note that such a resampler samples invalid particles even for the simplest case when $d=2$. Therefore, when using an unconstrained Gaussian distribution for the proposed protocol, some correction of states is generally required. The eigenvalue truncation and normalization used in conventional resamplers, and truncating Gaussian distributions as done in the proposed resampler are two ways to proceed. Compared to the unconstrained Gaussian resampler, TG i) always samples valid states for $d=2$ and ii) reduces the probability of sampling invalid states by removing the possibility of sampling states outside the unit-ball. In comparison to the conventional resamplers, the advantage of TG in terms of infidelities is slight. However, TG requires fewer computational resources. Since TG resampler is inherently sequential for each particle (although resampling of each particle is independent and therefore parallelizable), it scales as $O \left(d^{2}P \right)$ while conventional resampler scales as $O \left(d^{3} P \right)$ due to the requirement of eigendecomposition, where $P \gg d$ is the total number of particles. Moreover, it is interesting to note, that despite the unit-ball simplification, there are no evident drawbacks, and even states with high purity seldom require eigenvalue correction as the final step.

Fig.~\ref{fig:Noise} shows the robustness of our scheme in the presence of statistical noise which can be controlled by varying $N_{0}$, the number of measurements per iteration. We find that the asymptotic scaling of infidelity is independent of $N_{0}$ \cite{Fer:14:NJP}, \cite{MMM:21:PRL} for both pure and mixed qubit states. The initial advantage offered by smaller values of $N_{0}$ is due to the adaptive scheme. As the number of iterations increase, all curves in the inset of Fig.~\ref{fig:Noise} converge.

Moreover, we have also provided a quantum circuit that works hybridly with our particle filter. Although it estimates $\rho$ to a fair extent, it would be best if it is taken as only a proof of concept. The circuit in Fig.~\ref{fig:circuit} uses the simplest techniques to prepare $\rho$ and rotate measurement operators. More efficient circuits can be used that reduce the decoherence of the prepared state and in turn help us improve our estimates of the true state \cite{RCA:16:PRL}, \cite{AAR:18:IT}. Moreover, this technique applies a new measurement in each iteration. Although this requirement is fundamental for initial iterations, more attention can be afforded to its actual efficacy later on. If we can know that after certain $n$ iterations, QST estimates $\hat{\rho_{n}}$ with measurement configuration $k$ and after $n+1$ it estimates $\hat{\rho}_{n+1}$ with measurement configuration $k+1$ where the trace distance $\delta \left( \hat{\rho_{n}}, \hat{\rho}_{n+1} \right) \approx 0$, the computational overhead in calculating and changing to configuration $k+1$ can be avoided without significant difference to the scaling of our estimate.
\section{Conclusion}\label{sec:5}
We proposed an adaptive Bayesian QST technique for multiple qubits that changes the measurement basis in each iteration such that it seamlessly uses the prior as an effective first step. We have reported the numerical and experimental infidelities in our estimates and their uncertainties for arbitrary two-dimensional states. Furthermore, we have provided a comparison of infidelities with popular Bayesian particle filter methods used for QST and demonstrated our advantage in estimation over them. One prospective work can be to use the empirically derived prior to produce novel adaptive quantum tomography methods which take advantage of the maximum possible L1 error of the first approximated state. Maximum L1 error of the first estimate $\hat{\rho}$ is the maximum absolute distance between the corresponding elements of the Bloch vectors of $\rho$ and $\hat{\rho}$. An advantage of such a method is that it allows us to evaluate lower bounds of infidelity with respect to $N_{0}$ and total number of iterations.


\begin{acknowledgements}
We acknowledge use of the IBM Q for this work. The views expressed are those of the authors and do not reflect the official policy or position of IBM or the IMB Q team. This work was supported by the National Research Foundation of Korea (NRF) grant funded by the Korea government (MSIT) (No. 2019R1A2C2007037).
\end{acknowledgements}


\bibliographystyle{splncs}		

\end{document}